\documentclass{svproc}
\usepackage{graphicx}

\usepackage{url}

\begin{document}
\mainmatter              
\title{Probing Ultra-light Primordial Black Holes as a Dark Matter Candidate}
\author{Anupam Ray}
\institute{Tata Institute of Fundamental Research, Homi Bhabha Road, Mumbai, 400005, India\\
\email{anupam.ray@theory.tifr.res.in}}

\maketitle              

\begin{abstract}
Dark Matter (DM) is omnipresent in our universe. Despite its abundance, the microscopic identity of DM still remains a mystery. Primordial black holes
(PBHs), possibly formed via gravitational collapse of large density perturbations in the early universe, are one of
the earliest proposed and viable DM candidates. Recent studies indicate that PBHs can make up a
large or even entire fraction of the DM density for a wide range of masses. Here, we briefly review
the observational constraints on PBHs as DM, concentrating on the ultra-light mass window. Ultra-light
PBHs emit particles via Hawking radiation and can be probed by observing such Hawking evaporated
particles in various space as well as ground based detectors. We also outline how next-generation gamma ray
telescopes can set a stringent exclusion limit on ultra-light PBH DM 
by probing low energy photons.

\end{abstract}

\section{Introduction}
\label{sec:1} 
Cosmological observations provide unambiguous evidence of a non-relativistic, collision-less, and weakly interacting non-baryonic matter, commonly known as dark matter (DM), as a dominant component of the Universe~\cite{Aghanim:2018eyx}. Many well-motivated DM candidates have been proposed, and decades of experimental searches have been conducted to hunt for these mysterious DM particles, yet the microscopic identity remains
unknown. Primordial black holes (PBHs), possibly formed via gravitational collapse of large overdensities in the very early universe  are one of the earliest proposed and
viable DM candidates~\cite{Novikov,Hawking,Chapline}. 

PBHs have a wide range of masses depending on their time of formation, and can constitute a large or even entire fraction of the present day DM density in the mass range of $\sim 10^{17}-10^{23}$\,g. The idea of PBHs as DM has recently received a renewed attention with the detection of a BH merger by the LIGO-Virgo
collaboration and the subsequent proposals that the BH merger can be primordial in origin~\cite{Bird,Cleese,Sasaki}. Several techniques have been implemented in order to probe the PBH fraction of DM,  yields a multitude of observational constraints~\cite{Kohri,Carr,Green}. Here, we briefly review the current status of PBHs as DM, mostly concentrating on the ultra-light mass window.

\section{Ultra-light PBHs as Dark Matter}

\begin{figure}
	\centering
	\includegraphics[width=0.55\textwidth]{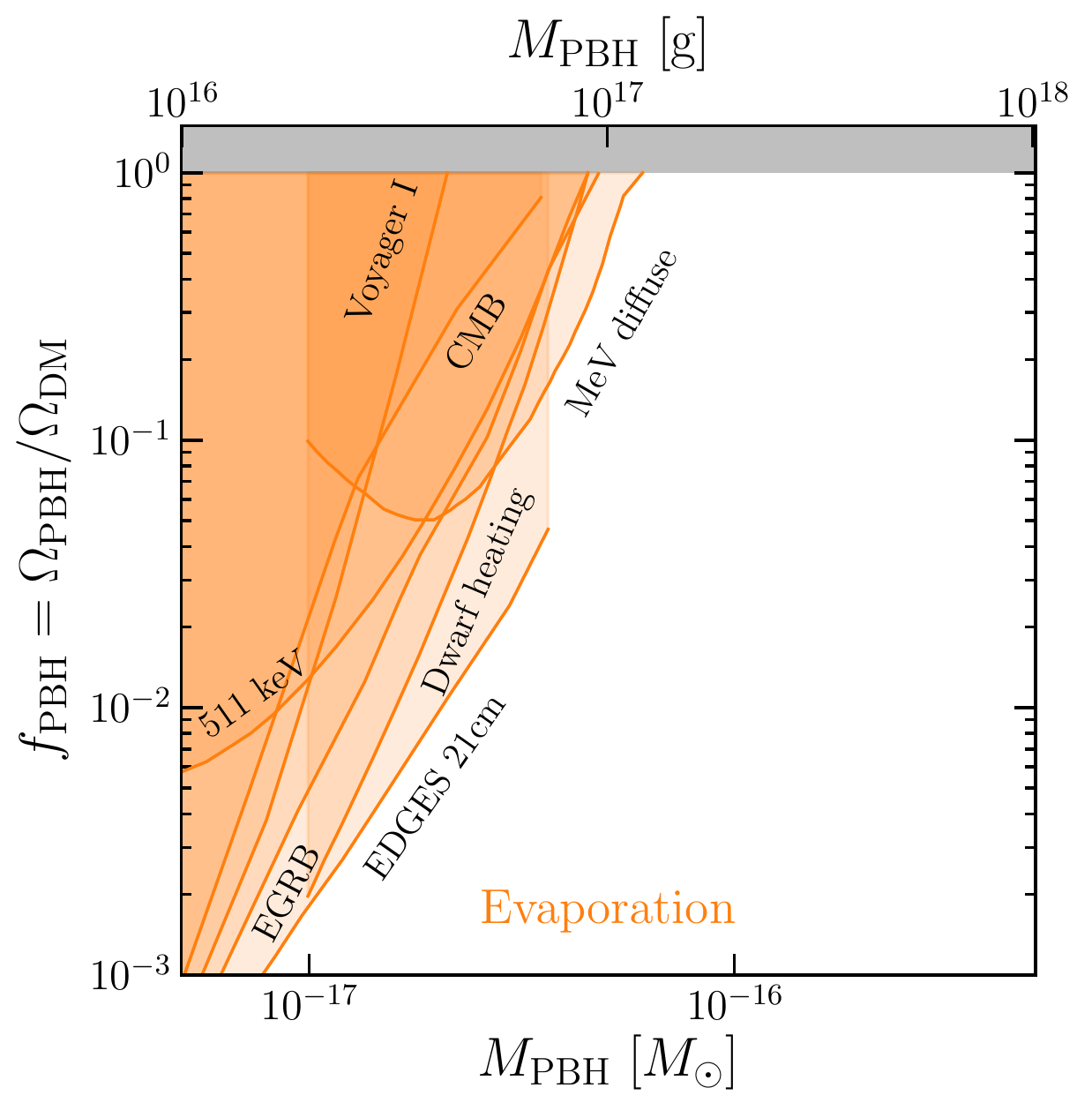}
	\caption{ Consolidated upper limit on the fraction of DM composed of ultra-light evaporating PBHs, $f_{\rm{PBH}}$, from non-observations of Hawking radiated particles in various space as well as ground based detectors.  Non-rotating PBHs with a monochromatic mass distribution is considered for this plot, and the figure is adapted from Ref.~\cite{Green}.
  }
\label{evap}
\end{figure}

PBHs evaporate via Hawking radiation and the evaporation timescale is proportional to the cube of their masses. The lifetime of  non-rotating (maximally rotating) PBHs lighter than $5\times10^{14}$\, g ($7\times10^{14}$\, g) is less than the age of our universe, hence, they cannot contribute to the present day DM density, setting the lower mass limit of ultra-light PBHs as DM~\cite{Page1,Page2}. Heavier PBHs i.e.\, PBHs with masses in between $5\times 10^{14}$\,g to $2\times10^{17}$\,g are typically probed via searching their evaporation products. Non-observation of such Hawking evaporated neutrinos~\cite{Dasgupta}, photons~\cite{Kohri2,Arbey,Ballesteros,Laha,Coogan}, and electrons/positrons~\cite{Dasgupta,Boudad,DeRocco,Laha2} provide the  leading exclusions on the PBH fraction of dark matter. Fig.\,\ref{evap} provides a consolidated view of the existing constraints for the ultra-light evaporating PBHs in the mass range of $10^{15}-10^{17}$\,g.

\emph{Neutrino derived exclusions}: Ref.~\cite{Dasgupta} shows that non-observation of $\mathcal O (10)$ MeV neutrinos in the diffuse supernovae neutrino background (DSNB) searches by an underground neutrino observatory Super-Kamiokande sets stringent constraints on ultra-light PBHs as DM. For a monochromatic mass distribution, it excludes non-rotating (maximally rotating) PBHs to form the sole component of DM upto $5\times10^{15}$\,g ($10^{16}$\,g), and for an extended mass distribution (e.g.\,log-normal mass distribution with a width of $\sigma=1.0$), it even excludes upto $\sim10^{17}$\,g.

\emph{Photon derived exclusions}: Observations of extra-Galactic gamma ray background (EGRB) via several space based telescopes such as COMPTEL, FERMI LAT, SMM exclude monochromatic, non-rotating  PBHs to form the sole component of DM up to $\sim10^{17}$\,g~\cite{Kohri2}.  The exclusion limits cover more mass window for rotating PBHs, and allowing maximum rotation  can extend the constraints upto $6\times10^{17}$\,g for a monochromatic mass distribution~\cite{Arbey}. Ref.~\cite{Ballesteros} shows that astrophysical modeling of the EGRB sources can significantly improve the photon derived exclusions.  Galactic Center gamma-ray measurements by the space based telescope INTEGRAL also provides stringent constraints on the PBH fraction of DM~\cite{Laha}. For a monochromatic mass distribution, it excludes non-rotating PBHs to form solitary component of DM upto $\sim 2 \times10^{17}$\,g, and for maximally rotating PBHs, it excludes upto $10^{18}$\,g. Ref.~\cite{Coogan} obtains a similar result by using the Galactic Center gamma-ray measurements by COMPTEL.

\emph{Electron/Positron derived exclusions}: Measurement of the $e^{\pm}$ flux by  the spacecraft Voyager 1 significantly constrains the fraction of DM composed of ultra-light PBHs~\cite{Boudad}.  Refs.~\cite{DeRocco,Laha2} show that measurement of the 511 keV gamma-ray line also provide stringent exclusions on ultra-light PBHs as DM. It excludes non-rotating, monochromatic PBHs to form the sole component of DM upto $\sim10^{17}$\,g.  However, the positron derived constraints crucially depend on various astrophysical uncertainties, such as unknown positron propagation distance as well as choice of DM density profiles, and can vary by almost an order of magnitude due to these uncertainties. Ref.~\cite{Dasgupta} extends the positron derived analysis for PBHs that have rotation, and shows that allowing maximal rotation along with extended mass distributions of the PBHs make the constraints more stringent, and can even extend them upto $\sim10^{19}$\,g. 

\emph{Photon derived projected exclusions}: Non-rotating, monochromatic PBHs in the mass range of $\sim10^{17} $– $10^{23}$\,g can make up the entirety of the present day  DM density as the existing exclusion limits from Hawking evaporation cease at $\sim 10^{17}$\,g  and the existing exclusion limits from optical microlensings start from $\sim 10^{23}$\,g. Ref.~\cite{Ray}
shows that measurement of the low energy ($\sim$ MeV) photons from the Galactic Center with the imminent telescopes such as AMEGO can probe this entirely unexplored mass window.  AMEGO can exclude non-rotating (maximally rotating) monochromatic PBHs as a sole component of DM upto $7\times10^{17}$\,g ($4\times10^{18}$)\,g by assuming no signal is present in the data. Ref.~\cite{Coogan} also obtains similar results by using a future telescope GECCO.

\emph{Summary}: Non-rotating monochromatic PBHs with masses $\leq 10^{17}$\,g can not form the solitary component of DM because of the non-observations of Hawking evaporated particles in various space as well as ground based detectors. For PBHs that have rotation, or which follow extended mass distributions, these exclusion limits probe more mass windows. However, non-rotating monochromatic PBHs in the mass range of $\sim10^{17} $– $10^{23}$\,g can form the entirety of the present day DM density as the exclusion limits in that mass window are invalidated by several recent studies.  Measurements of the low energy photons from the Galactic Center by upcoming soft gamma-ray telescopes such as AMEGO can close this mass window by almost an order of magnitude.

\end{document}